# Polarisation selective magnetic vortex dynamics and core reversal in rotating magnetic fields.

M. Curcic[1], B. Van Waeyenberge[1,2*], A. Vansteenkiste[2], M. Weigand[1], V. Sackmann[1], H. Stoll[1], M. Fähnle[1], T. Tyliszczak[3], G. Woltersdorf[4], C. H. Back[4] and G. Schütz[1]

[1.] *Max-Planck-Institut für Metallforschung, 70569 Stuttgart, Germany.*

[2.] *Department of Subatomic and Radiation Physics, Ghent University, 9000 Gent, Belgium.*

[3.] *Advanced Light Source, LBNL, 94720 Berkeley, CA, USA.*

[4.] *Institut für Experimentelle und Angewandte Physik, Universität Regensburg, 93040 Regensburg, Germany.*

\* Corresponding author: bartel.vanwaeyenberge@UGent.be

**A magnetic vortex occurs as an equilibrium configuration in thin ferromagnetic platelets of micron and sub-micron size[1,2] and is characterised by an in-plane curling magnetisation. At the centre, a magnetic singularity is avoided by an out-of-plane magnetisation core.[3,4] This core has a gyrotropic excitation mode, which corresponds to a circular motion of the vortex around its equilibrium position[5] with a rotation sense determined by the direction of the vortex core magnetisation, its polarisation. Unlike linear fields[6] or spin polarised currents,[7] an in-plane rotating field can selectively excite one of the polarisation states.[8,9] Here we report the observation of vortex dynamics in response to rotating magnetic fields, imaged with**

**time-resolved scanning X-ray microscopy.[10] We demonstrate that the rotating field only excites the gyrotropic mode if the rotation sense of the field coincides with the vortex gyration sense, and that such a field can selectively reverse the vortex polarisation.**

A detailed understanding of the static and dynamic properties of ferromagnetic thin film structures is crucial for their successful implementation in technological devices, like magnetic sensors or magnetic memories. Nano-scaled magnetic thin film structures like magnetic wires or platelets are interesting objects to study the fundamental dynamics of magnetic configurations like domain walls and vortices.

In thin film structures of soft magnetic materials, the stray field is minimised by forcing the magnetisation in the sample plane and forming inhomogeneous configurations like, e.g., domain walls. In micron and sub-micron sized structures of suitable thickness, this can result in a stable magnetic vortex configuration. The in-plane curling magnetisation in such elements turns out of the plane at the centre, avoiding a singularity and forming the vortex core. The vortex core has only a radius of about 5-15 nanometers,[4] but plays a key role in the magnetisation dynamics. The out-of-plane magnetisation of the vortex core (or 'polarisation', $p$) can point in two directions, 'up' or 'down' ($p=+1$ or $p=-1$, respectively). Together with the sense of the in-plane flux closure (or 'circulation', $c=+1(-1)$ for (counter) clockwise magnetisation), $p$ defines the magnetic configuration and it gives the vortex a defined handedness (right-handed for $cp=+1$, left-handed for $cp=-1$).

According to the equation of motion for magnetic structures, introduced by Thiele[11], and applied to vortex structures by Huber,[12] a moving vortex experiences an in-plane gyro force **F** perpendicular to its velocity and of which the sign is given by the core polarisation. This gives rise to a counterclockwise gyration for $p=+1$ and a clockwise gyration for $p=-1$. When a vortex is displaced from its equilibrium position, the restoring force from the demagnetising fields and the gyro force result in a spiral motion back into the equilibrium position This motion corresponds to the fundamental excitation mode of the vortex structure, with a frequency typically in the sub-GHz range and can be directly excited by an alternating magnetic field. This mode was first observed more than 20 years ago,[5] but the recent advances in magnetic microscopy have revealed unexpected details: it was discovered that the excitation of this mode can toggle the vortex core polarisation[13].

In-plane uniaxial excitations, like linear oscillating fields, do not break the symmetry of mirror-symmetric structures like squares or disks, and would result in symmetric dynamics for vortices of opposite polarisations. The resulting switching process should be symmetric too: the excitation simply toggles the polarisation state back and forth and thus the final state should depends on both the amplitude and the duration of excitation. This is different for in-plane rotating fields, which couple differently to the two polarisation states[14] and hence lead to an asymmetric switching behaviour.[8,9,15]

In this letter, we report on the experimental observation of the vortex core dynamics due to in-plane rotating magnetic fields. 500x500x50 nm$^3$ square Permalloy elements with a magnetic vortex configuration were excited with an in-plane rotating magnetic field ***B***($t$)=($B_0$cos($\omega t$),$B_0$sin($\omega t$),0)) at a frequency $|\omega/(2\pi)|$ close to the frequency of the gyrotropic eigenmode, and were investigated by time-resolved magnetic X-ray microscopy (see Figure 1 and the Methods section).

Figure 2 shows the result of the sequence of magnetic fields that was applied to one of the vortex structures. A clockwise (CW) rotating field was applied to a vortex down state (Figure 2, row a, p=-1). This rotation sense coincides with the gyrotropic rotation sense of the core and the core gyration can clearly be observed. When the amplitude of the rotating field is increased, the gyration amplitude increases (row b), until at an amplitude of 0.310 mT, the polarisation of the vortex core reverses and the gyration stops. (row c). Even when the amplitude was decreased again, the vortex appeared static. Then, the same procedure was repeated with a counter clockwise (CCW) rotating field to the vortex 'up' state and a CW gyration is observed (row d, e) until at an rf field amplitude of 0.345 mT where the vortex motion stops again (row f). This sequence was repeated many times and the response of the vortex always followed the same scheme.

These observations indicate that the core can be unidirectionally switched using rotating fields: an 'up' vortex and a 'down' vortex can be reversed by, respectively, a CCW and a CW rotating field of sufficient amplitude. When the core polarisation has switched, the rotating field can no longer excite the vortex to the threshold speed required for switching [16,17] unless the rotation sense of the field is reversed as well.

Our experimental results agree qualitatively with those of recent micromagnetic simulations[9] and with our own simulations (see Methods). A CCW rotating field with an amplitude exceeding 0.73 mT, could reverse an 'up' vortex. The same characteristic stages of the vortex core reversal as were first identified by Van Waeyenberge *et al.*[13] (i.e., via formation and annihilation of vortex-antivortex pairs) could be recognized. Figure 3 shows the vortex core trajectories extracted from this simulation with a 4 mT rotating field. We observe that the vortex is quickly accelerated counter clockwise by the applied field until the core polarisation switches. Now the vortex is spiralling back clockwise towards the centre of the structure. Here, the rotation sense changes again and only a small CCW motion remains. A similar motion without any switching event is observed when a CW rotating field of 4 mT is applied to the initial configuration. Only when the amplitude of a CW field is increased above 22 mT, the vortex core switches, but keeps switching back and forth. These simulations are consistent with the experimental observations if we consider that the small motion after switching is below the resolution of the microscope from the vortex core trajectories shown in figure 2. At the low excitation amplitude of 0.13 mT, the speed is about 120 m/s for both vortex polarisations. At the higher excitation amplitudes, just below the switching threshold (0.34 mT for the 'up' vortex and 0.30 mT for the 'down' vortex), the vortex speed is about 190 m/s. This speed is slightly lower than the speed threshold for switching of 320 m/s estimated by Guslienko[17] and of 330±15 m/s estimated from our own simulations. One observation was not reproduced by the simulations: the different switching thresholds for the two polarizations. In perfect samples, micromagnetic simulations show that the 'up' and 'down' vortex are energetically equal and the dynamics are symmetric when respectively excited by a CCW and CW rotating magnetic field. An asymmetry between the two core polarisations has been observed before in a detailed measurement of the trajectories of the two core states.[18] Such asymmetries can only be understood if the *p* symmetry is broken. It is tempting to assume that this is caused by an asymmetry of the sample geometry or the experimental conditions. However, so far no such asymmetry could be identified which would be strong enough to lead to the observed effect. However, a *p*-asymmetry for a vortex with given circulation may be the result of the appearance of a vector which carries the information on both the circulation and the polarization (e.g., an axial vector), in either the expression for the magnetic energy (then the asymmetry appears already for the ground state) or in the equation of motion for the atomic magnetic moments $M_i$ at sites *i* (then the asymmetry appears for dynamical properties like the ac switching field). An example is the axial vector $M_i$ x $M_j$ which enters the Dzyaloshinskii-Moriya (DM)[19,20] interaction in systems with broken structural inversion symmetry.[21] In our samples the inversion symmetry could be broken due to the unlike top and bottom interfaces of the Permalloy layer or due to residual strain from the thim film deposition.[22] It remains to be investigated by calculations with the density-functional electron theory whether for realistic assumptions on the structure of the sample the DM interaction is strong enough to explain the observed effect. In fact, the DM interaction would even predict a preferred handedness for a given sample with given structural properties. We performed similar experimental investigations as described above on a couple of different samples and found that the preferred handedness and the magnitude of the difference between the two switching fields depend on the sample, and sometimes the magnitude even changed while the sample was measured. Within the DM interpretation this would indicate different structural properties of the different samples as well as strain release due to heating during the course of measurement.

To conclude, we showed that the fundamental eigenmodes of the two vortex core polarisation states can be selectively excited by in-plane rotating magnetic fields. Such fields of sufficient strength can switch the vortex polarity in an unidirectional way. By changing the rotation sense of the field, vortices with either an 'up' or 'down' core magnetisation can be excited. The experimental results show an asymmetry in threshold amplitudes of the two polarisation states for switching and suggests a breaking of symmetry.

**Methods**

A set of 500x500x50 nm$^3$ square Permalloy (Py, Ni$_{80}$Fe$_{20}$) elements were patterned on top of two crossing copper striplines (see Figure 1), 150 nm thick and 10 μm wide. To ensure transparency for soft X-rays, the striplines and Py elements where constructed on a 100 nm thick Si$_3$N$_4$ membrane. An in-plane rotating magnetic field **B**(*t*)=(cos(ω*t*),sin(ω*t*),0) with frequency |ω/(2π)| was generated by running two oscillating currents with a 90° phase difference to both arms of the crossed striplines. Each stripline was driven through a balun to ensure equal currents in both ends of the stripline and to isolate the signal from the other arm. By changing the phase of one of the two signals by 180°, the rotation sense of the field can easily be changed from counterclockwise (CCW, ω>0) to clockwise (CW, ω<0). By switching off one of the signals, a linear alternating field can also be applied. Because the generated magnetic field is not completely homogeneous over the stripline, the rotating field will effectively be elliptical away from the centre of the cross. We have estimate this difference between the major and minor axes of the ellipse to be less then 3% for the investigated elements.

The time-dependent magnetisation distribution of the sample was imaged by a stroboscopic measurement technique, with a scanning transmission X-ray microscope at the Advanced Light Source (ALS, beamline 11.0.2)[23]. The contrast mechanism used for the imaging of the magnetic structure is the X-ray magnetic circular dichroism (XMCD) effect, namely the dependence of the absorption coefficient of circularly polarized monochromatic X-rays on the direction of the magnetisation in a ferromagnetic sample.[24] The elliptically polarized monochromatic X-rays provided by the undulator beamline are focused by a Fresnel zone plate to a spot of about 30 nm. The sample is scanned through the focal spot with a high-resolution scanning stage while the transmitted photon flux is measured. The data was recorded at the Ni L$_3$-absorption edge (852.7 eV), at which the XMCD effect gives a strong magnetic contrast.

For right (left) circularly polarised light, the absorbed X-ray intensity is (inverse) proportional to the projection of the magnetisation vector **M** on the photon propagation vector **k**. When the sample is placed perpendicular to the photon beam(see Figure 1), only the out-of-plane magnetisation contributes to the magnetic contrast, so the vortex core can be directly imaged and its polarisation can be derived. Alternatively, the sample normal was tilted by 30° with respect to the photon beam to image the in-plane magnetisation of the investigated Landau structures. In this geometry, the observed magnetic contrast is dominated by the contribution from the *x*-component of the in-plane closed flux magnetisation and the chirality of the vortex structure can be verified.

In order to record snapshots of the magnetic response in a stroboscopic fashion, the oscillating excitation field is synchronized with the probing X-ray flashes. This allows us to image the magnetisation with a time resolution near 70 ps, as given by the width of the X-ray flashes. An excitation frequency of 562.5 MHz was chosen, which is 9/8 of the photon flash repetition frequency and close to the gyrotropic resonance frequency. This allows us to simultaneously record the response at 8 different phases of the excitation by using an 8-channel acquisition.[25]

The OOMMF code[26] was used to perform 2-dimensional micromagnetic simulations. Using Py material parameters (saturation magnetisation $M_s=733 \times 10^3$ A/m, exchange stiffness $A=13 \times 10^{-12}$ J/m, crystalline anisotropy constant $K_1=0$, damping parameter alpha=0.01), the response of a 500 nm x 500nm x 50 nm element was simulated. The cell size in *x*- and *y*-direction was 2 nm, and in the *z*-direction 50 nm. The excitation frequency was set to 562 MHz, the gyrotropic resonance frequency of the simulated structure.

**Acknowledgements**
Cooperations with Aleksandar Puzic, Kang Wei Chou and Michael Hirscher are gratefully acknowledged. We would also like to thank Sabine Seiffert and Christian Wolter for the mechanical construction of the sample holder. A.V. acknowledges the financial support by The Institute for the promotion of Innovation by Science and Technology in Flanders (IWT Flanders). The Advanced Light Source is supported by the Director, Office of Science, Office of Basic Energy Sciences, of the US Department of Energy.

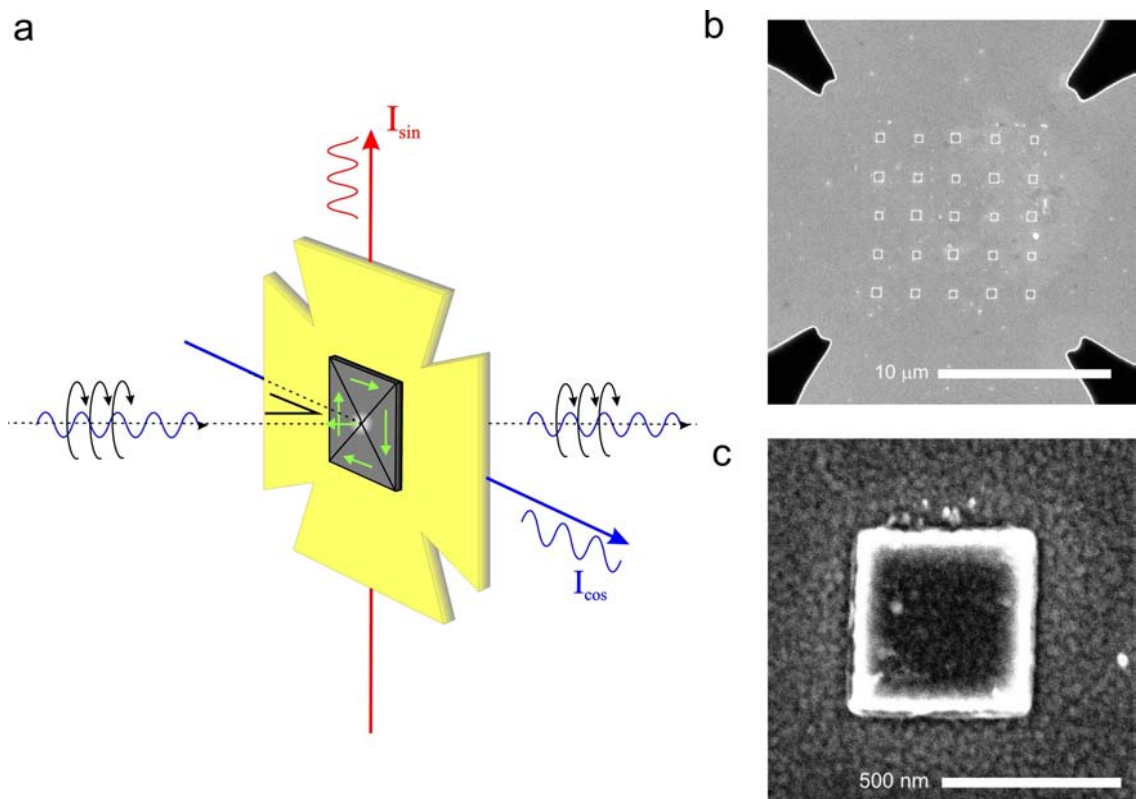

Figure 1: Schematic overview of the sample geometry. (a) The white spot reflects the magnetic contrast of the 'up' vortex core as given by the XMCD effect. The top inlay (b) shows a SEM micrograph of the crossed strip line with an array of Py elements. The bottom inlay (c) shows a SEM micrograph of the single 500 nm x 500 nm x 50 nm Py element at the centre of the array of which the X-ray absorption image are shown in figure 2.

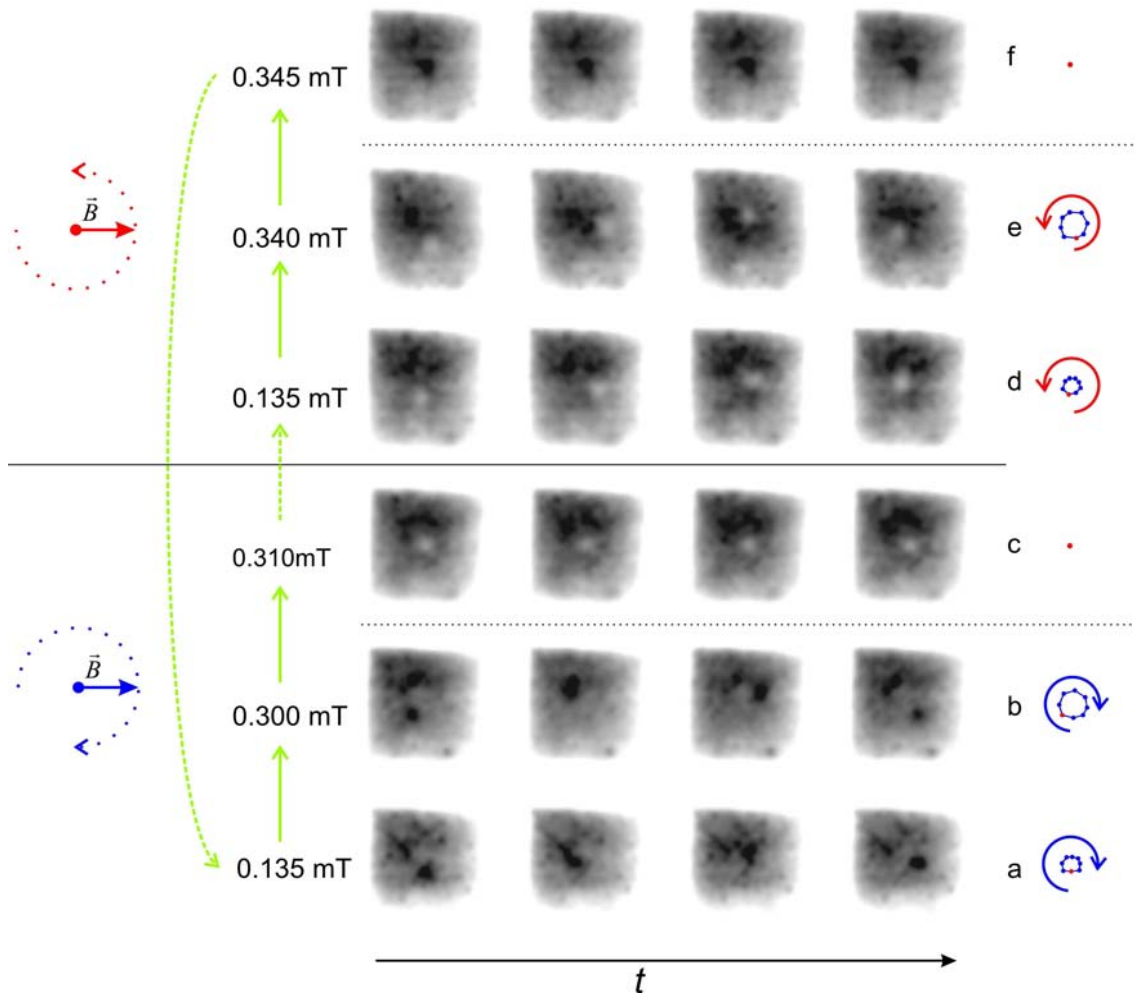

Figure 2: Time resolved imaging of the vortex core with right circularly polarised photons in a 500 x 500 x 50 nm$^3$ magnetic vortex structure excited by an in-plane rotating magnetic field. The out-of-plane magnetisation of the vortex core gives rise to a bright or dark spot, depending if the polarisation of the core is positive or negative, respectively. Four frames of each movie corresponding to a full period of the applied counter clockwise (rows a – c) and clockwise (rows d – f) rotating magnetic field. On the right hand side the corresponding vortex core trajectories are given, derived from all 8 acquired movie frames.

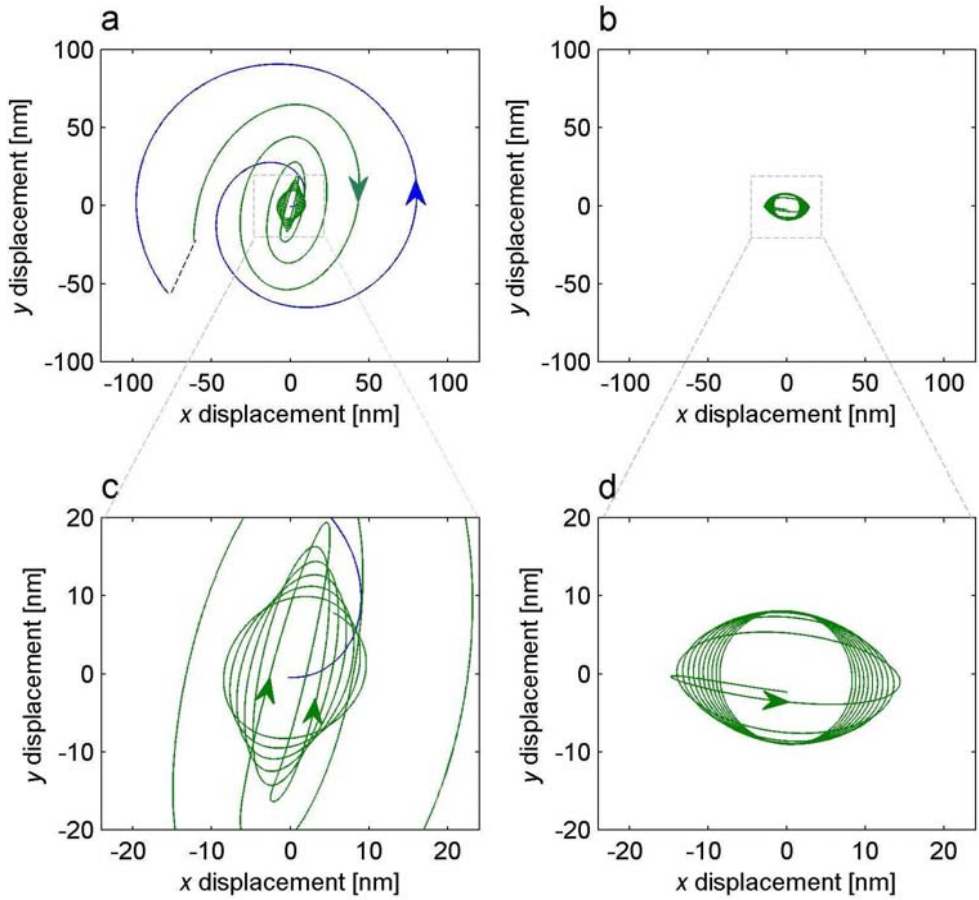

Figure 3: Trajectories from micromagnetic simulations, left: panel a and c, a vortex with core polarisation $p = +1$ excited by a CCW rotating field, panel b and d, a vortex with core polarisation $p = -1$ excited by a CCW rotating field. The colours indicate the vortex polarisation, green = down, blue = up.